\documentstyle[preprint,aps,epsf]{revtex}
\begin{document}
\def\gev{{\rm \,Ge\kern-0.125em V}}
\def\lo{\,{_{ out}\langle}}
\def\ro{\rangle_{out }\,}
\def\linn{\,{_{ in}\langle}}
\def\rinn{\rangle_{in }\,}
\def\lox{{_{  }\langle}}
\def\rox{\rangle_{  }}
\def\linnx{{_{  }\langle}}
\def\rinnx{\rangle_{  }}
\def\lout{\langle}
\def\lin{{\langle}}
\def\rin{\rangle}
\def\rout{\rangle}
\def\H{H_{I}}
\def\T{{\bf T}}
\def\inti{\int_{-\infty}}
\def\gtwid{\mathrel{\raise.3ex\hbox{$>$\kern-.75em\lower1ex\hbox{$\sim$}}}}
\def\ltwid{\mathrel{\raise.3ex\hbox{$<$\kern-.75em\lower1ex\hbox{$\sim$}}}}
\preprint{hep-ph/0110387}
\draft
\title{Baryon Number Violation in Particle Decays}
\author{
RATHIN ADHIKARI$^{(a)*}$
and RAGHAVAN RANGARAJAN$^{(b)\dagger}$}
\address{
$^{(a)}$ Jagadis Bose National Science Talent Search\\ 716, Jogendra Gardens, 
Kasba, Kolkota (Calcutta) 700 078, India\\ 
$^{(b)}$ Theoretical Physics Division, Physical Research Laboratory\\
Navrangpura, Ahmedabad 380 009, India\\
}
%\date{\today}
\maketitle
\begin{abstract}
\baselineskip 18pt

It has been argued in the past that in baryogenesis via out-of-equilibrium    
decays one must consider loop diagrams that contain more than one baryon 
number violating coupling.  In this note we argue that the
requirement with regard to baryon number violating couplings in
loop diagrams is that the interaction between the intermediate on-shell
particles and the final particles should correspond to a net
change in baryon number and that this can be satisfied even
if the loop diagram contains only one baryon number violating
coupling.  Put simply, we show that to create a baryon asymmetry
there should be net B violation
to the right of the `cut' in the loop diagram.
This is of relevance to 
some works involving the out-of-equilibrium decay scenario.

\vspace{1in}
* E-mail: rathin\_adhikari@yahoo.com
 
$\dagger$ E-mail: raghavan@prl.ernet.in
\end{abstract}
\pacs{}

\newpage
\baselineskip 18pt

\narrowtext

%%\section{}
%\section{Introduction}
%%\label{sec:intro}

It is well known that to obtain a baryon asymmetry in the out-of-equilibrium
baryon number
violating decays of heavy particles one must consider the interference between
tree level diagrams and higher order loop diagrams.  Furthermore, some particles
in the loop must be able to go on shell for the net asymmetry to be non-zero.
This is typically illustrated by drawing a `cut' 
through the lines representing particles that have gone on shell.
In the Appendix of Ref.~\cite{wn}, the authors had argued that a further
requirement is that one must consider loop diagrams that contain more than
one B violating coupling.  In this brief note we argue 
that the requirement with regards to B violating
couplings is that
the interactions on the right of the `cut' should correspond 
to a non-zero change
in the baryon number.  Furthermore, this can be satisfied
even if the loop diagram contains only
one B violating coupling and we refer the reader to such an example.

Consider a particle
$X$ and its antiparticle $\bar X$ 
each of which can decay to final states with different baryon number.
Let $f$ be a specific final state with baryon number
$B_F$ that $X$ decays to.  
Assuming CP violation, the partial decay rates for $X$ going to specific
final states and for $\bar X$ going to the corresponding final states 
can be
different. 
Therefore we now consider the amplitude $A(\bar X\rightarrow \bar f)$
for the decay of $\bar X$ to $\bar f$.  By the $CPT$ theorem,
\begin{equation}
A(\bar X\rightarrow \bar f)=A(f\rightarrow X) \, .
\end{equation}
Therefore
\footnote{
The issue of whether it is possible to have unstable particles in asymptotic
states \cite{veltman} is ignored in Ref.~\cite{wn} and by us.
}  
\begin{equation}
\lo \bar f| \bar X \rinn 
=\lo X|f\rinn\, .
\end{equation}
Inserting a complete set of
$in$ states 
\footnote{As in Ref.~\cite{wn}, we shall henceforth drop the subscript for the
$|X\rangle$ states as $|X\rinn=|X\ro$ for one particle states.}
\begin{eqnarray}
%\lin \bar f|S| \bar X \rin
\lox X|f\rinn 
%&=&\sum_g \lo X|g\ro \lo g|f\rinn\nonumber\\
&=&\sum_g \linnx X|g\ro\lo g|f\rinn
\end{eqnarray}
The sum over states above includes integration over momenta.
Then 
\begin{eqnarray}
\sum_{\bar f_{B_F}}|\lo \bar f| \bar X \rinnx |^2&=&
\sum_{f_{B_F}} \sum_g \linnx X|g\ro\lo g|f\rinn \lox X| f\rinn^*\nonumber\\
&=&\sum_{ f_{B_F}} \sum_g \linnx X|g\ro\lo g|f\rinn \linn f| X\rox\nonumber\\
&=&\sum_{ f} \sum_g \linnx X|g\ro\lo g|f\rinn \linn f| X\rox
-\sum_{ f_B\ne B_F} \sum_g \linnx X|g\ro\lo g|f\rinn \linn f| X\rox\nonumber\\
&=&\sum_{ f} \sum_{g_{B_F}} \linnx X|g\ro\lo g|f\rinn \linn f| X\rox\nonumber\\ 
&&+\sum_{ f} \sum_{g_B\ne B_F} \linnx X|g\ro\lo g|f\rinn \linn f| X\rox 
-\sum_{ f_B\ne B_F} \sum_g \linnx X|g\ro\lo g|f\rinn \linn f| X\rox\nonumber\\
&=& \sum_{g_{B_F}} \linnx X|g\ro\lo g| X\rox \nonumber\\
&&+\sum_{ f} \sum_{g_B\ne B_F} \linnx X|g\ro\lo g|f\rinn \linn f| X\rox 
-\sum_{ f_B\ne B_F} \sum_g \linnx X|g\ro\lo g|f\rinn \linn f| X\rox\nonumber\\
&=& \sum_{g_{B_F}} \linnx X|g\ro\lo g| X\rinnx\nonumber\\ 
&&+\sum_{ f_{B_F}} \sum_{g_B\ne B_F} \linnx X|g\ro\lo g|f\rinn \linn f| X\rox 
+\sum_{ f_{B}\ne B_F} \sum_{g_B\ne B_F} \linnx X|g\ro\lo g|f\rinn \linn f| X\rox
\nonumber\\
&& 
-\sum_{ f_B\ne B_F} \sum_{g_{B_F}} \linnx X|g\ro\lo g|f\rinn \linn f| X\rox
-\sum_{ f_B\ne B_F} \sum_{g_B\ne {B_F}} \linnx X|g\ro\lo g|f\rinn \linn f| X\rox
\nonumber\\
&=& \sum_{g_{B_F}} |\lo g| X\rinnx|^2 \nonumber\\
&&
+\sum_{ f_{B_F}} \sum_{g_B\ne B_F} \linnx X|g\ro\lo g|f\rinn \linn f| X\rox 
-\sum_{ f_B\ne B_F} \sum_{g_{B_F}} \linnx X|g\ro\lo g|f\rinn \linn f| X\rox
\nonumber\\
&=& \sum_{g_{B_F}} |\lo g| X\rinnx|^2 \nonumber\\
&&
+\sum_{ f_{B_F}} \sum_{g_B\ne B_F} \biggl [
\linnx X|g\ro\lo g|f\rinn \linn f| X\rox 
- \linnx X|f\ro\lo f|g\rinn \linn g| X\rox \biggr ]\nonumber\\
&=& \sum_{g_{B_F}} |\lo g| X\rinnx|^2 \nonumber\\
&&
+\sum_{ f_{B_F}} \sum_{g_B\ne B_F} \biggl [
A^*(X\rightarrow g) A(f\rightarrow g) A^*(f\rightarrow X)\nonumber\\
&&\hfil - A^*(X\rightarrow f) A(g\rightarrow f) A^*(g\rightarrow X)\biggr ]
\end{eqnarray}
where the sum over $f_{B_F} ({\bar f_{B_F}})$ 
is over all final states with a fixed baryon number,
$B_F (-B_F)$, and includes integration over momenta of the
$f (\bar f)$ states.  

The net decay rate, $\bar\Gamma(-B_F)$, for $\bar X$ to all final states 
with baryon number $-B_F$ is proportional to the expression on the 
left-hand side above,
while the net decay rate, $\Gamma(B_F)$, for $X$ to all final states
with baryon number $B_F$ is proportional to the first term on the right-hand
side above.

For a 2-body decay scenario, the difference term, i.e.,
the second term on the right-hand side above, is 0 to $O(\lambda^2)$,
where $\lambda$ is any coupling in the theory.  At $O(\lambda^4)$,
the difference term can be rewritten as 
\begin{eqnarray}
&&\sum_{ f_{B_F}} \sum_{g_B\ne B_F} \biggl [
A_c^*(X\rightarrow g) A_c^*(g\rightarrow f) A_c(X\rightarrow f)
+ A_c^*(X\rightarrow f) A_c(g\rightarrow f) A_c(X\rightarrow g)
\biggr ]\nonumber\\
&=& 2 Re\sum_{ f_{B_F}} \sum_{g_B\ne B_F} \biggl [
A_c(X\rightarrow g) A_c(g\rightarrow f) A_c^*(X\rightarrow f)
\biggr ]\, ,     
\end{eqnarray}
where $A_c$ is the connected (tree level) amplitude and
we have used $A_c^*(a\rightarrow b)=-A_c(b\rightarrow a)$
which is valid for diagrams in which no internal particles go on shell.

Thus to obtain a difference in $\Gamma(B_F)$ and $\bar\Gamma(-B_F)$ one
requires that 
there exist intermediate states $g$ such that 
the transition between the states $g$ and the
final states $f$ involves a net change in baryon number.  In other
words, the requirement for an asymmetry is that there must exist diagrams
such that the process to the right
of the `cut' should violate baryon number and the net asymmetry is then
proportional to the amplitudes associated with {\it these} diagrams.  

If the process $X\rightarrow g$, where $B_g\ne B_F$,
involves a B violating coupling then the total one loop amplitude
for $X\rightarrow f$ must involve more than one B violating coupling for it 
to contribute to the net asymmetry.  In
this case, the statement that the one loop diagram must involve more than
one B violating coupling to obtain an asymmetry, as argued in Ref. \cite{wn}, 
holds.  This
is shown in fig. 1.  In fact, the insertion of internal states $g$ in 
Ref. \cite{wn} implicitly assumes that the process $X\rightarrow g$
involves B violation.  However,
an asymmetry between $\Gamma(B_F)$ and $\bar\Gamma(-B_F)$ may be achieved
even if the one loop amplitude for $X\rightarrow f$ involves only one
B violating coupling.  If, for example, $X$ carries no baryon number and
$B_F=1$ then the one loop amplitude for the process $X\rightarrow f$
in which the states $g$ carry no baryon number can
involve only one B violating coupling (see fig. 2) and yet satisfy the
requirement for an asymmetry discussed above.  
This is the case in Ref. \cite{claudsonetal},
albeit for a 3-body decay scenario.  
(Our conclusions can be extended to the asymmetry from
3 body decays.)

If the processes $g\rightarrow f$ involve one or
more than one B violating coupling, as in fig. 3, 
the total contribution of the corresponding loop diagrams to the
net asymmetry will still be 0 if $B_g=B_F$.  
Note that the asymmetry is not 0 for individual decay channels when
$B_g=B_F$.  
But summing over all intermediate and final states gives zero net
asymmetry for processes in which
the intermediate states and
the final states have the same baryon number.  This result is relevant for
the calculations of the asymmetry in, for example, Ref. \cite{mas,as}.

\acknowledgements

One of us (R.A.) is supported by DST, India.  R.A. and R.R. thank the
Physical Research Laboratory, Ahmedabad and the Institute for Mathematical
Sciences, Chennai respectively for
hospitality during  this study.

\oddsidemargin=-6pt
\topmargin=-1.25in
\textwidth=6.5in
\textheight=9.4in
\voffset=0.75in

\newpage
\begin{figure}[htb]
\mbox{}
\vskip 1.8in\relax\noindent\hskip -0.1in\relax
\includegraphics{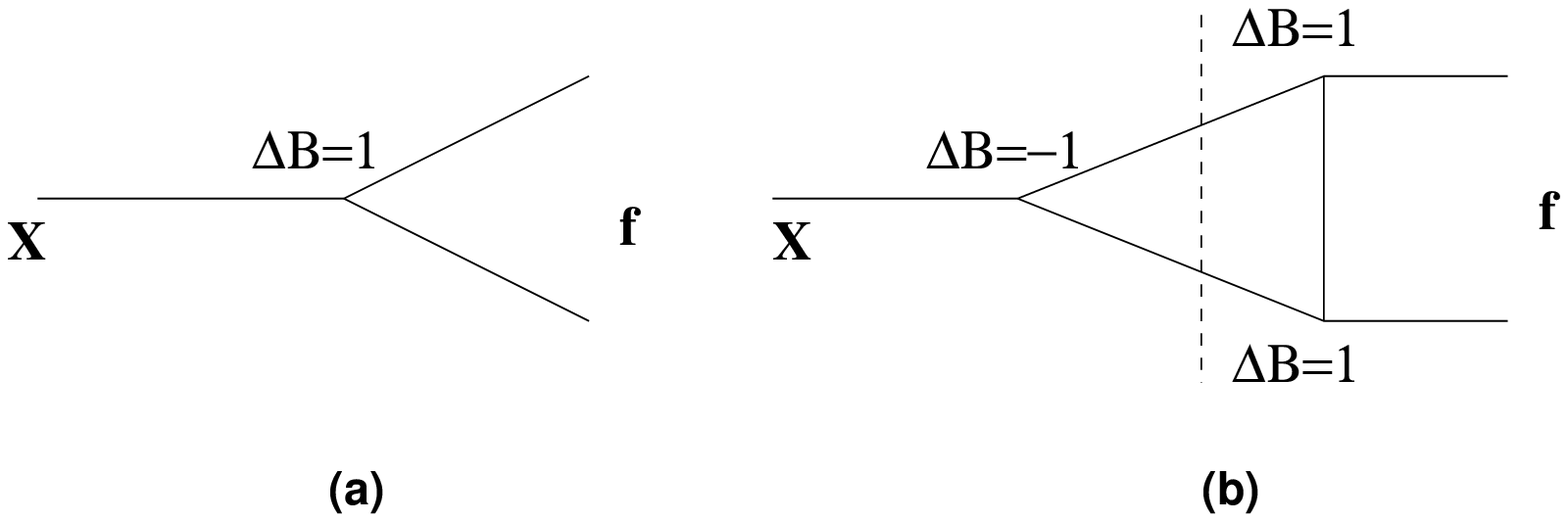} \vskip .25in
\caption{   Tree level and one loop diagrams for $X\rightarrow f$ and net
$\Delta B=1$.  The loop diagram has $B_g\ne B_F$ and involves more than one
$B$-violating coupling.} 
\end{figure}
\begin{figure}[htb]
\mbox{}
\vskip 2.8in\relax\noindent\hskip 0.8in\relax
\includegraphics{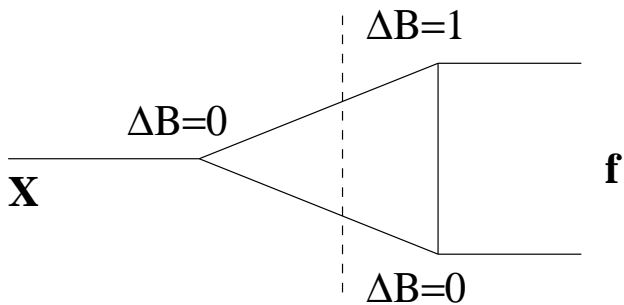} \vskip .25in
\caption{   One loop  diagram  with $B_g\ne B_F$ and only one $B$-violating  
coupling.}
\end{figure}
\newpage
\begin{figure}[hbt]
\mbox{}
\vskip 1.8in\relax\noindent\hskip 0.8in\relax
\includegraphics{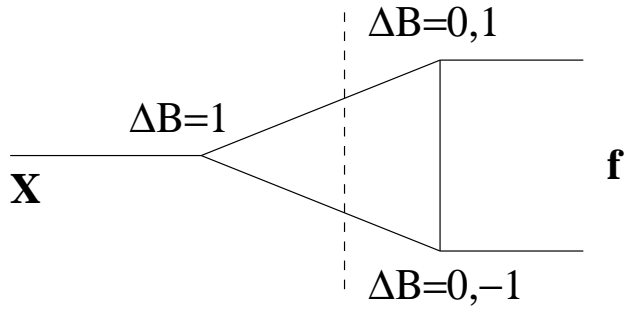} \vskip .25in
\caption{   One loop  diagram  with $B_g=B_F$ and with one and more than one 
$B$-violating  couplings.}
\end{figure}

\end{document}